\documentclass[12pt,dvips]{article}
\usepackage{color,pstricks}
\usepackage{amsmath,amssymb,epsfig}
\usepackage{psfig,epsfig}
\usepackage{array}
\psdraft
\def\chain#1#2#3{{\relax\hbox{
        \raise 40pt\vtop{
               \begin{tabbing}
                       $#1{}$\=$#2$\\
                       \>$\; |$\\ [-9pt]
                       \>$\ \rightarrow #3$\\
               \end{tabbing}           }}}}

\begin{document}
\newcommand{\ra}        {\mbox{$\rightarrow$}}
\newcommand{\bc}        {\begin{center}}
\newcommand{\ec}        {\end{center}}
\newcommand{\be}        {\begin{equation}}
\newcommand{\ee}        {\end{equation}}
\newcommand{\gam}       {\mbox{$\gamma$}}
\newcommand{\piz}       {\mbox{$\pi^0$}}
\newcommand{\pip}       {\mbox{$\pi^+$}}
\newcommand{\pim}       {\mbox{$\pi^-$}}
\newcommand{\etg}       {\mbox{$\eta$}}
\newcommand{\etp}       {\mbox{$\eta^{\prime}$}}
\newcommand{\omg}       {\mbox{$\omega$}}
\newcommand{\rh}        {\mbox{$\rho$}}
\newcommand{\pbp}       {\mbox{$\bar{p}p$}}
\newcommand{\ssb}       {\mbox{$\bar{s}s$}}
\newcommand{\NNb}       {\mbox{$\overline{N}N$}}
\newcommand{\pbar}       {\mbox{$\bar{p}$}}

\definecolor{lightyellow}{cmyk}{0,0,0.5,0}
\definecolor{lightred}{rgb}{1,0.5,0.5}
\definecolor{lightgreen}{rgb}{0.5,1,0.5}
\definecolor{lightblue}{rgb}{0.5,0.5,1}
\definecolor{darkgreen}{rgb}{0,0.5,0}
\definecolor{darkcyan}{cmyk}{1,0.3,0.3,0.3}
\definecolor{darkblue}{rgb}{0.5,0.5,1}
\definecolor{lightbrown}{rgb}{0.7,0.3,0.3}
\definecolor{darkbrown}{rgb}{0.5,0,0}
\definecolor{bluegreen}{rgb}{0,0.5,0.5}
\def\chain#1#2#3{{\relax\hbox{
        \raise 40pt\vtop{
               \begin{tabbing}
                       $#1{}$\=$#2$\\
                       \>$\; |$\\ [-9pt]
                       \>$\ \rightarrow #3$\\
               \end{tabbing}           }}}}
\definecolor{fill1}{rgb}{0.5,0.0,0.0}
\definecolor{shadow1}{rgb}{0.25,0.0,0.0}
\definecolor{fill2}{rgb}{0.0,0.5,0.0}
\definecolor{shadow2}{rgb}{0.0,0.25,0.0}
\definecolor{fill3}{rgb}{0.34,0.0,0.72}
\definecolor{shadow3}{rgb}{0.17,0.0,0.36}
\definecolor{fill_gl}{rgb}{0.9,0.9,0.9}
\definecolor{shadow_gl}{rgb}{0.45,0.45,0.45}
\newcommand{\nonett}[9]
{
\setlength{\unitlength}{1mm}
\begin{picture}(150.00,90.00)
\put(10.00,45.00){\vector(1,0){70.00}}
\put(45.00,10.00){\vector(0,1){70.00}}
\put(110.00,45.00){\vector(1,0){30.00}}
\put(125.00,30.00){\vector(0,1){30.00}}
\put(82.50,42.50){\makebox(5.00,5.00){$I_3$}}
\put(142.50,42.50){\makebox(5.00,5.00){$I_3$}}
\put(127.50,60.00){\makebox(5.00,5.00){S\quad\ Singlet }}
\put(47.50,80.00){\makebox(5.00,5.00){S\quad\ Octet }}
\put(45.00,45.00){\circle*{2.00}}
\put(70.00,45.00){\circle*{2.00}}
\put(20.00,45.00){\circle*{2.00}}
\put(32.50,70.00){\circle*{2.00}}
\put(57.50,70.00){\circle*{2.00}}
\put(32.50,20.00){\circle*{2.00}}
\put(57.50,20.00){\circle*{2.00}}
\put(125.00,45.00){\circle*{2.00}}
\put(45.00,45.00){\circle{0.00}}
\put(45.00,45.00){\circle{5.00}}
\put(32.50,70.00){\line(1,0){25.00}}
\put(57.50,70.00){\line(1,-2){12.50}}
\put(70.00,45.00){\line(-1,-2){12.50}}
\put(57.50,20.00){\line(-1,0){25.00}}
\put(32.50,20.00){\line(-1,2){12.50}}
\put(20.00,45.00){\line(1,2){12.50}}
\put(60.00,70.00){\makebox(15.00,5.00)[l]{#2}}
\put(15.00,70.00){\makebox(15.00,5.00)[r]{#1}}
\put(15.00,15.00){\makebox(15.00,5.00)[r]{#3}}
\put(6.75,37.50){\makebox(13.25,5.00)[r]{#5}}
\put(60.00,15.00){\makebox(15.00,5.00)[l]{#4}}
\put(70.00,37.50){\makebox(13.75,5.00)[l]{#6}}
\put(47.50,37.50){\makebox(12.50,5.00)[l]{#7}}
\put(47.50,47.50){\makebox(12.50,5.00)[l]{#8}}
\put(127.50,47.50){\makebox(12.50,5.00)[l]{#9}}
\end{picture}
}

%
%
%
%
%
\newcommand{\Ra}{$\bf\Rightarrow$}
\renewcommand{\bottomfraction}{0.9}
\renewcommand{\topfraction}{0.9}
\renewcommand{\textfraction}{0.1}
%
\title{Do parity doublets in the baryon spectrum reflect restoration
of chiral symmetry\,? }

\vskip 30mm
\author{E. Klempt
\\
Helmholtz-Institut f\"ur Strahlen -- und Kernphysik, \\
Universit\"at Bonn,\\
D--53115 Bonn }
\maketitle
{\it Abstract: We discuss the mass spectrum of highly-excited
nucleon and $\Delta^*$ resonances. The spectrum exhibits parity
doublets, pairs of resonances of identical total angular momentum
J but of opposite parity. It has been proposed that the parity
doublets evidence restoration of chiral symmetry at large
baryon excitation energies. We compare this conjecture with the
possibility that high-mass states are organized into $(L,S)$-multiplets
with defined intrinsic quark spins and orbital angular momenta.
The latter interpretation results in a better
description of the data. There is however a small
trend possibly indicating the onset of 
chiral symmetry restoration.
}
\bc
revised 2003, March 10 \\
to be published in Physics Letters B
\ec
\section{Introduction}
The observation of parity doublets in high-mass excitations of the
nucleon and of the $\Delta$ has stimulated a discussion if this
effect signals restoration of chiral symmetry
\nocite{Cohen:2002st,Glozman:2002kq,Glozman:2002cp,Cohen:2001gb,Glozman:1999tk}
\cite{Cohen:2002st}-\cite{ Glozman:1999tk}. At high masses,
resonances can be grouped into doublets of states having the same
total angular momentum $J$ but opposite parities. At lower masses,
chiral symmetry is broken, and the mass of the chiral partner of
the nucleon, the N(1535)$S_{11}$, differs from the nucleon mass
substantially. In this letter, we will denote resonances like the
N(1535)$S_{11}$ as N$_{1/2^-}(1535)$ where spin and
parity are given explicitely.
\par
Chiral symmetry allows for separate parity doublets in the nucleon and
the $\Delta$ sector even though it supports also a higher
symmetry in which N$^*$ and $\Delta^*$ resonances of a given $J$ and
opposite in parity are all degenerate in mass. Data seem to
support this higher symmetry. This interpretation is  however not
uncontested: in a relativistic quark model with instanton induced
forces, nucleonic parity doublets arise naturally
\nocite{Loring:2001bp,Loring:2001kv,Loring:2001kx,Loring:2001ky}
\cite{Loring:2001bp}-\cite{Loring:2001ky}. However, none of the present
quark model calculations reproduces the parity doublets in the
$\Delta^*$ mass spectrum 
\nocite{Capstick:bm,Glozman:1997ag,Glozman:1997fs}
\cite{Capstick:bm}-\cite{Glozman:1997fs}. 
The interpretation of the parity 
doublets as evidence for chiral symmetry restoration seems
thus unavoidable. 
\par
In this letter we suggest a different interpretation
of parity doublets. We show that parity doublets develop
naturally when spin orbit forces are neglected. The symmetry
leading to the occurence of parity doublets is thus identified as
absence of spin-orbit forces.
\par
Table \ref{chiral}, adapted from Cohen and Glozman
\cite{Cohen:2001gb}, shows N$^*$ and
$\Delta^*$ masses above 1.9\,GeV, for states with positive and negative
parity. In many cases, the effect of parity doubling is striking:
states with identical $J$ but opposite parity often have very similar
masses. This does of course not imply that chiral symmetry restoration 
is the reason for the occurence of parity doublets. 
\par
Consider e.g. the first six $\Delta$ states in Table~\ref{chiral}
with $J=1/2, 3/2$ and 5/2, and with positive and negative parities
\cite{weinheimer}.
The masses are clearly degenerate; they form three parity doublets.  
The $\Delta_{7/2^+}(1950)$ and the $\Delta_{7/2^-}(2200)$ should also
form a parity doublet but the $\Delta_{7/2^+}(1950)$ has a mass which is
very close to the other three positive-parity resonances; the four
positive-parity resonances rather seem to belong to a spin quartet
of states with intrinsic orbital angular momentum $L=2$ and intrinsic spin $S=3/2$
coupling to $J=1/2,..,7/2$. 
The question arises
if the parity doublets are really due to restoration of chiral
symmetry or if the parity doublets reflect a 
symmetry of the underlying quark dynamics. 
\par
Cohen and Glozman \cite{Cohen:2001gb} emphasize that the scheme 
requires the existence of a $\Delta_{11/2^-}$ and a N$_{11/2^+}$ 
at about 2500 MeV, of a $\Delta_{13/2^-}$ and a N$_{13/2^-}$ at 
2750 MeV, and of three additional states at 2950 MeV. The existence of
these states is a compelling prediction of chiral symmetry
restoration. Experimental searches for these states are being carried
out at ELSA in Bonn \cite{crede}. Also at Jlab \cite{Napolitano:1993kf}, 
GRAAL \cite{Bocquet:2001ny}, and at Spring8 \cite{Zegers:2003ux} the
high-mass baryon spectrum is studied.   
\par
\vspace*{-1mm}
\begin{table}[h!]
\renewcommand{\arraystretch}{1.5}
\begin{center}
\begin{tabular}{|c|ccc|ccc|}
\hline
$J = \frac{1}{2}$ & 1& N$_{1/2^+}(2100)$ &
                       N$_{1/2^-}(2090)$ &
                    a& $\Delta_{1/2^+}(1910)$&
                       $\Delta_{1/2^-}(1900)$\\
$J = \frac{3}{2}$ & 2&N$_{3/2^+}(1900)$ &
                    N$_{3/2^-}(2080)$ &
                    b&  $\Delta_{3/2^+}(1920)$&
                    $\Delta_{3/2^-}(1940)$\\
$J = \frac{5}{2}$ & 3& N$_{5/2^+}(2000)$ &
                    N$_{5/2^-}(2200)$ &
                    c&  $\Delta_{5/2^+}(1905)$ &
                    $\Delta_{5/2^-}(1930)$ \\
$J = \frac{7}{2}$ & 4& N$_{7/2^+}(1990)$ &
                    N$_{7/2^-}(2190)$ &
                    d&  $\Delta_{7/2^+}(1950)$ &
                    $\Delta_{7/2^-}(2200)$ \\
$J = \frac{9}{2}$ & 5& N$_{9/2^+}(2220)$ &
                    N$_{9/2^-}(2250)$ &
                    e&  $\Delta_{9/2^+}(2300)$ &
                    $\Delta_{9/2^-}(2400) $ \\
$J = \frac{11}{2}$& 6& \bf N$\bf_{11/2^+}$  &
                    N$_{11/2^-}(2600)$ &
                    f&  $\Delta_{11/2^+}(2420)$&
                    $\bf\Delta_{11/2^-}$ (*) \\
$J = \frac{13}{2}$& 7& N$_{13/2^+}(2700)$ &
                    \bf N$\bf_{13/2^-}$  &
                    g&  $\bf\Delta_{13/2^+}$ &
                    $\Delta_{13/2^-}(2750)$ \\
$J = \frac{15}{2}$& 8& \bf N$\bf_{15/2^+}$    &
                     \bf N$\bf_{15/2^-}$   &
                    h&  $\Delta_{15/2^+}(2950)$&
                    $\bf\Delta_{15/2^-}$ (*)  \\
\hline
\end{tabular}
\vspace*{-1mm}
\renewcommand{\arraystretch}{1.0}
\caption{Parity doublets of N$^*$ and $\Delta^*$ resonances
of high mass, after \protect\cite{Cohen:2001gb}. The states in
boldface are predicted to have the same mass as their chiral
partner  when chiral symmetry is restored in the high-mass excitation
spectrum of baryon resonances. We suggest that the states marked
with a (*) should have considerably higher masses than their chiral
partners while the other three states in boldface should be
degenerate in mass with  corresponding states of opposite parity.
}
\label{chiral}
\vspace*{-12mm}
\end{center}
\end{table}
\section{N$^*$ and $\Delta^*$ resonances}
The discussion of which resonances one should expect, and at which masses,
seems to require an understanding of how three valence quarks interact
to form baryons and baryon resonances. This we do not have. Instead,
we emphasize regularities in the mass spectra which can be used to
identify leading quantum numbers.
\par
A baryon resonance can be characterized by its flavor structure, by its
spin $J$ and parity $P$. In addition, there are quantum numbers which are
not directly accessible: the total spin $\vec{J}$ can be decomposed into
its orbital and spin angular momentum; the total orbital angular momentum
$\vec{L}$ is a sum of two orbital angular momenta $\vec{l}_{\rho}$ and
$\vec{l}_{\lambda}$ of the two oscillators allowed in a three-body system,
$\vec{S}$ the sum of the three quark spins. In a relativistic situation,
$\vec{l}_{\rho}$, $\vec{l}_{\lambda}$, $\vec{L}$, and $\vec{S}$
are not observable. Further, a flavor-octet resonance
may have a symmetric or mixed-symmetry spacial wave function, and the 
spacial wave functions can have $n_{\rho}$ and $n_{\lambda}$
nodes, the baryon could be radially excited. 
The multitude of dynamical degrees of freedom leads to 
a rich spectrum. This is the problem of the {\it missing resonances}: 
quark models predict a much larger number of states than observed. 
\par
An alternative was proposed by Lichtenberg who suggested that
baryons should be considered as quark-diquark excitations
where two quarks are frozen into a quasi-stable subsystem
\cite{Anselmino:1992vg}.
This possibility was never scrutinized in a dynamical model; however,
resonances like the N$_{7/2^+}(1990)$ and $\Lambda_{7/2^+}(2020)$
are not easily accomodated in a diquark model. 
\par
We conjecture that the solution of the missing resonances might be
found in a different interpretation of diquark configurations. 
Quark models expand the wave functions into harmonic oscillator 
wave functions $\left|(l_{\rho},n_{\rho});(l_{\lambda},n_{\lambda})\right>$. 
It is plausible that baryon resonances are formed with one 
oscillator excited in the scattering process. Since it is not known
which one is hit, there is a coherent superposition of 
$\left|(l_{\rho},n_{\rho})\neq 0;(l_{\lambda},n_{\lambda})=0\right>$ and 
$\left|(l_{\rho},n_{\rho})=0;(l_{\lambda},n_{\lambda})\neq 0\right>$ 
wave functions. No baryon is excited to a
$\left|(l_{\rho},n_{\rho})\neq 0;(l_{\lambda},n_{\lambda})\neq 0\right>$
component of a wave function, at the moment the resonance is formed. 
For these states, the initial $l_{\rho}$ or $l_{\lambda}$ (only
one is non-zero) can be identified with the total orbital angular
momentum $L$ and the initial $n_{\rho}$ or $n_{\lambda}$ with 
$N=n_{\rho}+n_{\lambda}$ (which we define as radial quantum number).
The wave functions constructed in this way are in general no energy
eigenfunctions but should form a wave packet of energy eigenfunctions
with a defined phase of the rotation or vibration. 
\par
This constraint leads to a large reduction in the number of expected 
states. We leave open the question if the 'missing' states do not
exist or if they decouple from the $\pi$N system. Since most of the 
N$^*$ and $\Delta^*$ resonances were found in $\pi$N elastic 
scattering, they could have escaped detection so far. They should 
uncover themselves in photoproduction experiments of complex final
states \cite{Capstick:2000qj} which allow to study cascades of 
high-mass resonances. Two-oscillator excitations could be 
populated via pion emission from a high-mass resonance. 
\par
We now show that the leading quantum numbers, $L,S,J,N$ 
of the known N$^*$ and $\Delta^*$ resonances
can be identified in most cases, and that mixing between different
internal configurations is small. This is an old observation stressed,
e.g., by Feynman, Pakvasa and Tuan \cite{Feynman:aj}.
\par
Table \ref{next} shows all known N$^*$ and $\Delta^*$ resonances
except the 1-star $\Delta_{1/2^+}(1750)$ and $\Delta_{5/2^+}(2000)$.
The ground states N and $\Delta$ are known to be members of a
SU(6) 56-plet which decomposes into a spin-1/2 octet and 
a spin-3/2 decuplet with $L=0$. Likely, there is a small contribution
of $L=2$ in the wave function \cite{Tiator:2000iy} but this effect does not prevent
us from identifying $L=0$ as leading component. In any case, 
the spatial wave function of these
ground-state baryons is symmetric, and their spin-flavor wave function
must be symmetric, too. The antisymmetry of the
wave function w.r.t. the exchange of two quarks as required by
the Pauli principle is guaranteed by the three colors.
\par
In the first two rows of Table \ref{next} 
there are two series' of states having the same
quantum numbers as the ground state baryons, with mass square differences
of $a\sim 1.1$\,GeV$^2$. The Roper N$_{1/2^+}$(1440) and
the analogous state $\Delta_{3/2^+}(1600)$  are supposed to be
first radial excitations of
the respective ground states; the
N$_{1/2^+}$(1710) and N$_{1/2^+}$(2100) the second and third radial
excitation. The $\Delta_{3/2^+}(1920)$ could be a radial excitation
even though the assignment of intrinsic orbital angular momentum $L=2$
and quark spin $S=3/2$ is possible as well and perhaps more likely. 
Also the  N$_{1/2^+}$(2100) could belong to a quartet of states
with $L=2$ and $S=3/2$, yet its mass is rather high in comparison
to the other positive parity N$^*$ states assigned to $L=2$. We
prefer to reserve this entry for the N$_{1/2^+}(1986)$ proposed
by the SAPHIR collaboration \cite{Plotzke:ua}. 
\clearpage
\begin{table}[h!]
\bc
\renewcommand{\arraystretch}{1.3}
\hspace*{-12mm}\begin{small}
\begin{tabular}{|cccc|cccc|c|}
\hline
$D$&$S$&$L$&$N$&& &  &  & Mass (\ref{mass})  \\
\hline
\hline
56&1/2&0&0,1,2,3&N$_{1/2^+}$(939)
&N$_{1/2^+}$(1440)&N$_{1/2^+}$(1710)&$^1$N$_{1/2^+}$(2100) &939 MeV  \\
&3/2&0&0,1,2,3&$\Delta_{3/2^+}$(1232)
&$\Delta_{3/2^+}(1600)$ &$\Delta_{3/2^+}$(1920)& & 1232 MeV \\
\hline
\hline
70&$1/2$&$1$&$0$&& N$_{1/2^-}(1535)$&N$_{3/2^-}(1520)$&    & 1530 MeV  \\
&$3/2$&$1$&$0$&&N$_{1/2^-}$(1650)& N$_{3/2^-}(1700)$&N$_{5/2^-}(1675)$&1631 MeV\\
&$1/2$&$1$&$0$&& $\Delta_{1/2^-}(1620)$ &  $\Delta_{3/2^-}(1700)$
&& 1631 MeV \\
\hline
56&$1/2$&$1$&$1$&& N$_{1/2^-}$&N$_{3/2^-}$&    & 1779 MeV  \\
&$3/2$&$1$&$1$&& $^a\Delta_{1/2^-}(1900)$ &  $^b\Delta_{3/2^-}(1940)$
& $^c\Delta_{5/2^-}(1930)$ & 1950 MeV \\
\hline
70&$1/2$&$1$&$1$&& $^1$N$_{1/2^-}(2090)$&$^2$N$_{3/2^-}(2080)$&    & 2151 MeV  \\
&$3/2$&$1$&$1$&&N$_{1/2^-}$& N$_{3/2^-}$&N$_{5/2^-}$&2223 MeV\\
&$1/2$&$1$&$1$&& $\Delta_{1/2^-}(2150)$ &  $\Delta_{3/2^-}$
&& 2223 MeV \\
\hline
\hline
56&$1/2$&$2$&$0$&& N$_{3/2^+}(1720)$&N$_{5/2^+}(1680)$&    & 1779 MeV  \\
&$3/2$&$2$&$0$&$^a\Delta_{1/2^+}(1910)$ & $^b\Delta_{3/2^+}(1920)$ &
 $^c\Delta_{5/2^+}(1905)$ & $^d\Delta_{7/2^+}(1950)$ &  1950 MeV \\
\hline
70&$1/2$&$2$&$0$&& N$_{3/2^+}$&N$_{5/2^+}$&    & 1866 MeV  \\
&$3/2$&$2$&$0$&N$_{1/2^+}$ & $^2$N$_{3/2^+}(1900)$ &
 $^3$N$_{5/2^+}(2000)$ & $^4$N$_{7/2^+}(1990)$& 1950 MeV\\
&$1/2$&$2$&$0$&& $\Delta_{3/2^+}$ &  $\Delta_{5/2^+}$
&& 1950 MeV \\
\hline
\hline
70&$1/2$&$3$&$0$&& $^3$N$_{5/2^-}(2200)$&$^4$N$_{7/2^-}(2190)$&
    & 2151 MeV  \\
&$3/2$&$3$&$0$&N$_{3/2^-}$& N$_{5/2^-}$&N$_{7/2^-}$&
$^5$N$_{9/2^-}(2250)$ & 2223 MeV  \\
&$1/2$&$3$&$0$&&$\Delta_{5/2^-}$&
$^d\Delta_{7/2^-}(2200)$ & & 2223 MeV   \\
\hline
56&$1/2$&$3$&$1$&& N$_{5/2^-}$&N$_{7/2^-}$&
    & 2334 MeV  \\
&$3/2$&$3$&$1$&$\Delta_{3/2^-}$&$\Delta_{5/2^-}(2350)$&
$\Delta_{7/2^-}$ & $^e\Delta_{9/2^-}(2400)$ & 2467 MeV   \\
\hline
\hline
56&$1/2$&$4$&$0$&& N$_{7/2^+}$&$^5$N$_{9/2^+}(2220)$ &    & 2334 MeV  \\
&$3/2$&$4$&$0$&$\Delta_{5/2^+}$& $\Delta_{7/2^+}(2390)$ &
$^e$$\Delta_{9/2^+}(2300)$ & $^f\Delta_{11/2^+}(2420)$ & 2467 MeV \\
\hline
\hline
70&$1/2$&$5$&$0$&& N$_{9/2^-}$&$^6$N$_{11/2^-}(2600)$&     & 2629 MeV  \\
56&$3/2$&$5$&$1$&$\Delta_{7/2^-}$ &$\Delta_{9/2^-}$
&$^f\Delta_{11/2^-}$ & $^g\Delta_{13/2^-}(2750)$ & 2893 MeV \\
\hline
\hline
56&$1/2$&$6$&$0$&& $^6$N$_{11/2^+}$&$^7$N$_{13/2^+}(2700)$ &    & 2781 MeV  \\
&$3/2$&$6$&$0$&$\Delta_{9/2^+}$& $\Delta_{11/2^+}$ &
$^g\Delta_{13/2^+}$ & $^h\Delta_{15/2^+}(2950)$ & 2893 MeV \\
\hline
\hline
70&$1/2$&$7$&$0$&&$^7$N$_{13/2^-}$&$^8$N$_{15/2^-}$&     & 3032 MeV  \\
56&$3/2$&$7$&$1$&$\Delta_{11/2^-}$ &$\Delta_{13/2^-}$
&$^h\Delta_{15/2^-}$ & $^i\Delta_{17/2^-}$ & 3264 MeV \\
\hline
\hline
56&$1/2$&$8$&$0$&& $^8$N$_{15/2^+}$& N$_{17/2^+}$ &    & 3165 MeV  \\
&$3/2$&$8$&$0$&$\Delta_{13/2^+}$& $\Delta_{15/2^+}$ &
$^i\Delta_{17/2^+}$ & $\Delta_{19/2^+}$ & 3264 MeV \\
\hline
\end{tabular}\end{small}
\renewcommand{\arraystretch}{1.0}
\caption{Nucleon and $\Delta^*$ resonances and the quantum numbers
assigned to them. More resonances are expected from quark models
than shown here (see text for a discussion). {\protect \\ }
$D$ is the dimensionality of the SU(6)
representation, $S, L$ are intrinsic spin and orbital
angular momenta assigned to a given resonance; $N$ represents a
radial excitation quantum number. The
masses in the right-hand column are calculated using eq. 
(\protect\ref{mass}). Pairs of nucleon resonances marked
$^{1},^{2},^{3},..$ and $\Delta$ resonances $^{a},^{b},^{c},..$ are
interpreted as parity doublets in \protect\cite{Cohen:2001gb}. 
}
\label{next}
\ec
\end{table}
\clearpage
\par
In many cases, quantum numbers can be assigned to groups of states
on the basis of an evident multiplet structure.
The low-mass negative parity resonances  with $L=1$ 
cannot have a completely symmetric spatial wave function, hence they
cannot be assigned to a 56-plet and must be in a SU(6) 70-plet. 
The latter decomposes into
a N$^*$ spin doublet, a N$^*$ spin quartet and a $\Delta^*$ spin
doublet, in 
accordance with the experimental findings. These
states are listed in rows 3-5 in Table \ref{next}. 
\par
In line 7, we find 
a triplet of negative-parity $\Delta^*$ resonances at about 1930
MeV. We are tempted to assign $L=1, S=3/2$ to these states; however
spin $S=3/2$ and isospin $I=3/2$  require a symmetric spatial wave
function. This can only be achieved if not only the angular momentum
is excited (to $L=1$), also the radial wave function needs to
have a node. The only way to avoid this conclusion would be to
assign $L=3, S=1/2$ to the $\Delta_{5/2^-}(1930)$  and $L=1, S=1/2$
to the $\Delta_{1/2^-}(1900)$ and $\Delta_{3/2^-}(1940)$. We prefer to
consider these three states as a triplet. The gap
in mass square to the negative-parity doublet is $\sim 1.1$\,GeV$^2$
and we assign one unit of radial excitation energy $(N = 1)$ to these
states. 
\par
We may expect e.g. also resonances in a SU(6) 70-plet with
$(N=1,L=1,S=1/2)$. The 70-plet would contain a N$^*$ spin doublet
$1/2^-,3/2^-$ at 1866\,MeV, a N$^*$ spin quartet $1/2^-,3/2^-,5/2^-$
at 1950\,MeV and a $\Delta^*$ doublet $1/2^-,3/2^-$ also at 1950\,MeV.
There are no entries for these states in the Review of Particle 
Properties even though there are two resonances proposed by SAPHIR, a
N$_{1/2^-}(1897)$ \cite{Plotzke:ua} observed in the  N$\eta^{\prime}$
decay mode and a N$_{3/2^-}$(1895) \cite{Bennhold:1997mg} decaying 
into K$^+\Lambda$. These are good
candidates for the $(N=1,L=1,S=1/2)$ multiplet.
\par
In rows 8-10 we list further negative-parity resonances. Their
assignment as $L=1, N=2$ states is an educated guess. 
\par
Positive-parity baryons with $L=2$ are possible as  
$\frac{1}{\sqrt 2}\left|(2,0);(0,0)\pm (0,0);(2,0)\right>$ 
configurations
building a 56-plet and a 70-plet. The next rows (11-12) list a doublet 
of N$^*$'s and a quartet of $\Delta^*$'s belonging to the 56-plet. 
The N$^*$ quartet at 1950 MeV (row 14) is part of the 70-plet. All
these resonances could have partners with radial excitation $N$ but 
no candidates are known. (Except perhaps the  $\Delta_{5/2^+}(2000)$
for which two mass values, 1752 and 2200\,MeV, are listed by the Particle 
Data Group. The larger value would allow a $L=2,N=1$ assignment.)    
\par
For $L=3$ and $L=4$ we should expect a 
repetition of the pattern observed for $L=1$ and for $L=2$. Indeed,
the known states can be mapped onto the predicted pattern. 
\par
The quantum numbers of high-mass resonances can best be identified
when they are 'stretched' states, with their spin and orbital angular
momentum aligned. Their observation can be used to assign quantum
numbers to states where only one resonance of a spin multiplet is
observed. For $L=4$ there is no state which would need to be assigned
to a 70-plet. In particular, there is no N$_{11/2^+}$. For large
excitation energies, the largest total angular momenta J in a given
mass range is often given by $J=L+S$ with
$S=1/2$ for N$^*$ and $S=3/2$ for $\Delta^*$: 
spin and isospin are locked. 
\par
The most straightforward assignment for nucleon resonances in the
mass range above 2.5\,GeV is $(L,S) = (5,1/2)$ for the
N$_{11/2^-}(2600)$, and $(L,S) = (6,1/2)$ for the N$_{13/2^+}(2700)$. 
To the $\Delta_{13/2^-}$(2750) we assign $(L=5,S=3/2)$ 
and $N=1$ since for 
$N=0$, a one-oscillator excitation to $L=5$ cannot be fully symmetric.
The  $\Delta_{15/2^+}$(2950) should have $(L=6,S=3/2)$. These two states
are expected here to have the same mass. This expectation is not
really supported by the data but also not falsified, due to the large
experimental errors. 
\par

\section{Baryon masses}

The regularity of the excitation energies suggests a baryon
mass formula \cite{Klempt:2002vp}
which is discussed in this section. The mass formula
reproduces with good $\chi^2$ the masses of all but one baryons with
known spin and parity.
The baryon mass formula reads
\begin{equation}
\label{mass}
{\rm  M^2 = M_{\Delta}^2 + \frac{n_s}{3}\cdot  \left(M_{\Omega}^2 -
M_{\Delta}^2\right)}   +
a \cdot (L + N) -  I_{sym}\cdot 
{\rm \left(M_{\Delta}^2 - M_{\rm N}^2\right)}.
\end{equation}
\noindent
$\rm M_{N}, M_{\Delta}, M_{\Omega}$ are input parameters taken from 
\cite{Hagiwara:fs},  
$a = 1.142$\,GeV$^2$ is the Regge slope determined from the series of
light (isoscalar and isovector) mesons with quantum numbers
$J^{\rm PC} = 1^{--}, 2^{++}, 3^{--}, 4^{++}, 5^{--}, 6^{++}$.
There is no adjustable parameter in the mass formula.
\par
The first two terms define the offset masses of Regge trajectories
with $n_s$ strange quarks in the baryon. Regge trajectories are
usually drawn as functions of $J$. They can, however, also
be drawn as functions of $L$. The squared masses then increase 
linearly with $L$, with good consistency.  
A motivation for this dependence was given by
Nambu \cite{Nambu:1978bd}. Note that the physical picture behind the mass
formula is radically different from present quark models for baryon resonances.
Here, the baryonic mass-gain with $L$ is assigned to an increasing mass of the
flux tube connecting (nearly massless) quarks. In quark models, the mass
gain with $L$ is due to an increase of kinetic and potential energy of the
constituent quarks. 
\par
N is the radial excitation quantum number. There are 17 cases in which
baryon resonances are observed which are higher in mass but have the
same quantum numbers as a lower-mass state (see Table II in
\cite{Klempt:2002vp}) the Roper N$_{1/2^+}$(1440) being the best known
example. The spacings in mass square are nearly the same as those for
consecutive values of $L$. These facts require the $L+N$ dependence
of the baryon masses while $L+2N$ gives the harmonic-oscillator band. 
This observation has also been made by Bijker, Iachello and Leviatan
\cite{Bijker:yr,Bijker:2000gq}. They proposed a baryon 
mass formula which is based on a spectrum-generating algebra. 
The Hamiltonian is bilinear in six vector boson operators
constructed for the two oscillators, plus one scalar boson
operator. Excitations of $n_{\rho},n_{\lambda}$ are described 
as phonon vibrational excitations; calculated masses reproduce well
experimental values.
\par
The total angular momentum $J$ does not enter the formula. The
spin orbit or $\vec{L}\cdot\vec{S}$ coupling is supposed to
vanish or to be small. 
\par
The spin $S$ enters only through the last symmetry
term which is defined to reproduce the N-$\Delta$ mass difference.
It acts only on octet and singlet particles
having spin 1/2; N$^*$'s with spin 3/2 and $\Delta^*$'s
are predicted to be degenerate in mass.
$\rm I_{sym}$ is the fraction of the harmonic-oscillator 
wave function (normalized to the nucleon wave function) which is
antisymmetric in spin and flavor. It is given by

\renewcommand{\arraystretch}{1.3}
\qquad\qquad\qquad \begin{tabular}{lrccc}
$\rm I_{sym}  = $&1& for $S=1/2$ & octet particles in a 56-plet;& \\
$\rm I_{sym}  = $&1/2& for $S=1/2$ & octet particles in a 70-plet;&
\qquad\qquad\qquad\quad  \hfill (2) \\
$\rm I_{sym}  = $&3/2& for $S=1/2$ & singlet particles; &\\
$\rm I_{sym}  = $&  0& otherwise.&
\end{tabular}
\label{sym}
\renewcommand{\arraystretch}{1.3}

\noindent
Instantons and anti-instantons
induce interactions in quark pairs when they are
antisymmetric in both, in spin and in flavor. The data require, through the term
$\rm I_{sym}$, a mass contribution proportional to this fraction in the wave
function. It is this peculiar pattern of (2) which leads us to conclude
that deviations from the leading Regge trajectory originate from
instanton-induced interactions. In particular
the N-$\Delta$ mass
splitting is thus assigned to instanton-induced interactions and not to
magnetic spin-spin interactions due to one-gluon exchange. The
numerical agreement between predicted and observed baryon
masses is quite good. For 44 the N$^*$ and $\Delta^*$ resonances 
listed in \cite{Hagiwara:fs} the
$\chi^2$ is 40. With the same errors, the one-gluon exchange model 
results in $\chi^2=82$ calculated for the 32 resonances for which a
mass is given in \cite{Capstick:bm}; the one-boson exchange
model \cite{Glozman:1997ag} 
yields a $\chi^2$ of 8 but uses only the 14 resonances below
1700 MeV. The number of parameters used for the mass formula is 4, 
in the one-gluon-exchange model 10 and in the one-boson exchange
model 5. 
\par
It may be useful to exploit the predictive power of the mass formula
also for some states which are not related to the question of parity
doublets. We have included in Table~\ref{next} baryon resonances which are
unobserved so far, and masses predicted by eq.~(\ref{mass}). 
In particular a negative-parity doublet at 1779 MeV and a
positive-parity doublet at 1866 MeV is expected. 
Not listed are resonances with a $(L,S,N)$ assignment for which no
state is known. In the subsequent discussion we use only the
lowest and second lowest mass baryon in a given partial wave. Thus
uncertainties due to the problem of missing resonances are mostly
avoided.

\section{Chiral parity doublets versus SU(6) multiplets}
\par
The mass formula predicts parity doublets, either of identical
or of approximately equal masses. The origin of the mass doublets
is different for N$^*$ and  $\Delta^*$ resonances. 
We begin with a discussion of $\Delta^*$ resonances.
\par
The three $\Delta^*$ resonances $\Delta_{5/2^{-}}(1930)$,
$\Delta_{9/2^{-}}(2400)$, and $\Delta_{13/2^{-}}(2750)$
are unlikely to have intrinsic $L=3,5,7$, respectively, but rather
$L=1,3,5$. The $\Delta_{5/2^{-}}(1930)$ is nearly degenerate in mass
with two other negative-parity states, the $\Delta_{9/2^{-}}(2400)$
with the $\Delta_{5/2^{-}}(2350)$, 
suggesting that they all belong to a spin quartet
(see Table \ref{next}), that they  have $S=3/2$. Hence their wave
function is symmetric in spin and in flavor. The Pauli principle
now requires a symmetric spatial wave function but negative
parity states can have a symmetric wave function only when they
are also excited radially.
\par
\begin{figure*}[h!]
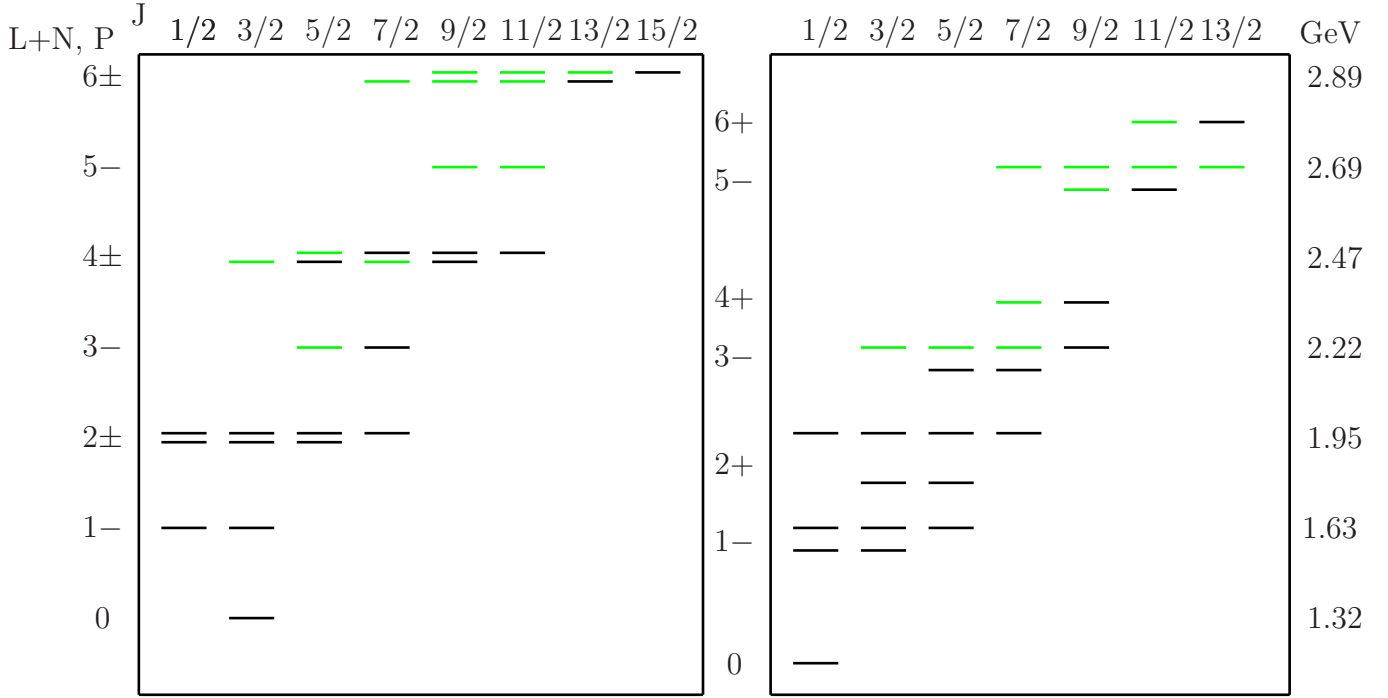

\vskip 40mm
\psset{xunit=.6cm,yunit=.6cm,linearc=.2}
\pspicture(-1,1)(19,5)
\psline[linewidth=1pt](13,-3.7)(13,10.5)
\psline[linewidth=1pt](0.5,-3.7)(13,-3.7)
\psline[linewidth=1pt](0.5,10.5)(13,10.5)
\psline[linewidth=1pt](0.5,-3.7)(0.5,10.5)
\psline[linewidth=1pt,linecolor=green](7,10.1)(8,10.1)
\psline[linewidth=1pt,linecolor=green](8.5,10.1)(9.5,10.1)
\psline[linewidth=1pt,linecolor=green](10,10.1)(11,10.1)
\psline[linewidth=1pt](11.5,10.1)(12.5,10.1)
\psline[linewidth=1pt,linecolor=green](5.5,9.9)(6.5,9.9)
\psline[linewidth=1pt,linecolor=green](7,9.9)(8,9.9)
\psline[linewidth=1pt,linecolor=green](8.5,9.9)(9.5,9.9)
\psline[linewidth=1pt](10,9.9)(11,9.9)
\psline[linewidth=1pt,linecolor=green](7,8.0)(8,8.0)
\psline[linewidth=1pt,linecolor=green](8.5,8.0)(9.5,8.0)

\psline[linewidth=1pt,linecolor=green](4,6.1)(5,6.1)
\psline[linewidth=1pt](5.5,6.1)(6.5,6.1)
\psline[linewidth=1pt](7,6.1)(8,6.1)
\psline[linewidth=1pt](8.5,6.1)(9.5,6.1)
\psline[linewidth=1pt,linecolor=green](2.5,5.9)(3.5,5.9)
\psline[linewidth=1pt](4,5.9)(5,5.9)
\psline[linewidth=1pt,linecolor=green](5.5,5.9)(6.5,5.9)
\psline[linewidth=1pt](7,5.9)(8,5.9)
\psline[linewidth=1pt,linecolor=green](4,4.0)(5,4.0)
\psline[linewidth=1pt](5.5,4.0)(6.5,4.0)

\psline[linewidth=1pt](1,2.1)(2,2.1)
\psline[linewidth=1pt](2.5,2.1)(3.5,2.1)
\psline[linewidth=1pt](4,2.1)(5,2.1)
\psline[linewidth=1pt](5.5,2.1)(6.5,2.1)
\psline[linewidth=1pt](1,1.9)(2,1.9)
\psline[linewidth=1pt](2.5,1.9)(3.5,1.9)
\psline[linewidth=1pt](4,1.9)(5,1.9)
\psline[linewidth=1pt](1,0)(2,0)
\psline[linewidth=1pt](2.5,0)(3.5,0)
\psline[linewidth=1pt](2.5,-2)(3.5,-2)

\rput[cl](1.7,11){1/2}\rput[cl](0.5,11.4){J}
\rput[cl](1.7,11){1/2}\rput[cl](3.2,11){3/2}
\rput[cl](4.7,11){5/2}\rput[cl](6.2,11){7/2}
\rput[cl](7.7,11){9/2}\rput[cl](9.2,11){11/2}
\rput[cl](10.7,11){13/2}\rput[cl](12.2,11){15/2}

\rput[cl](-1.1,10.8){L+N, P }
\rput[cl](-0.3,10){6$\pm$}
\rput[cl](-0.3,8){5$-$}
\rput[cl](-0.3,6){4$\pm$}
\rput[cl](-0.3,4){3$-$}
\rput[cl](-0.3,2){2$\pm$}
\rput[cl](-0.3,0){1$-$}
\rput[cl](-0.3,-2){0}

\rput(14,0){
\psline[linewidth=1pt](12,-3.7)(12,10.5)
\psline[linewidth=1pt](0.5,-3.7)(12,-3.7)
\psline[linewidth=1pt](0.5,10.5)(12,10.5)
\psline[linewidth=1pt](0.5,-3.7)(0.5,10.5)

\psline[linewidth=1pt,linecolor=green](8.5,9)(9.5,9)
\psline[linewidth=1pt](10,9)(11,9)

\psline[linewidth=1pt,linecolor=green](5.5,8.0)(6.5,8.0)  
\psline[linewidth=1pt,linecolor=green](7,8.0)(8,8.0)  
\psline[linewidth=1pt,linecolor=green](8.5,8.0)(9.5,8.0)              
\psline[linewidth=1pt,linecolor=green](10,8.0)(11,8.0)  

\psline[linewidth=1pt,linecolor=green](7,7.5)(8,7.5) 
\psline[linewidth=1pt](8.5,7.5)(9.5,7.5)             

\psline[linewidth=1pt,linecolor=green](5.5,5)(6.5,5) 
\psline[linewidth=1pt](7,5)(8,5)                     

\psline[linewidth=1pt,linecolor=green](2.5,4.0)(3.5,4.0)              
\psline[linewidth=1pt,linecolor=green](4,4.0)(5,4.0)  
\psline[linewidth=1pt,linecolor=green](5.5,4.0)(6.5,4.0)  
\psline[linewidth=1pt](7,4.0)(8,4.0)  

\psline[linewidth=1pt](4,3.5)(5,3.5)      
\psline[linewidth=1pt](5.5,3.5)(6.5,3.5)  

\psline[linewidth=1pt](1,2.1)(2,2.1)       
\psline[linewidth=1pt](2.5,2.1)(3.5,2.1)   
\psline[linewidth=1pt](4,2.1)(5,2.1)       
\psline[linewidth=1pt](5.5,2.1)(6.5,2.1)   

\psline[linewidth=1pt](2.5,1)(3.5,1)       
\psline[linewidth=1pt](4,1)(5,1)           
\psline[linewidth=1pt](1,-0.5)(2,-0.5)     
\psline[linewidth=1pt](2.5,-0.5)(3.5,-0.5) 
\psline[linewidth=1pt](1,-3)(2,-3)         

\rput[cl](1.7,11){1/2}\rput[cl](3.2,11){3/2}
\rput[cl](4.7,11){5/2}\rput[cl](6.2,11){7/2}
\psline[linewidth=1pt](1,0)(2,0)
\psline[linewidth=1pt](2.5,0)(3.5,0)
\psline[linewidth=1pt](4,0)(5,0)
\rput[cl](7.7,11){9/2}\rput[cl](9.2,11){11/2}
\rput[cl](10.7,11){13/2}
\rput[cl](12.9,11){GeV}

\rput[cl](-0.3,9){6$+$}
\rput[cl](-0.3,7.7){5$-$}
\rput[cl](-0.3,5.1){4$+$}
\rput[cl](-0.3,3.8){3$-$}
\rput[cl](-0.3,1.4){2$+$}
\rput[cl](-0.3,-0.3){1$-$}
\rput[cl](-0.3,-3){0}

\rput[cl](13,10){2.89}
\rput[cl](13,8){2.69}
\rput[cl](13,6){2.47}
\rput[cl](13,4){2.22}
\rput[cl](13,2){1.95}
\rput[cl](13,0){1.63 }
\rput[cl](13,-2){1.32}
}
\endpspicture
\vspace*{30mm}
\caption{Schematic diagram of the energy levels of $\Delta^*$ (left)
and N$^*$ (right) resonances. The vertical axis is linear in
squared baryon masses, mass values are given on the right axis.
In case of mass degenerate states, negative-parity states 
are drawn below those with positive parity. 
Observed states are denoted 
by dark lines, expected ones by grey lines.  }
\label{n-delta}
\end{figure*}

Figure \ref{n-delta} (left) shows the multiplet structure of  
$\Delta^*$  resonances. According to eq. (\ref{mass}) the masses
depend on $L+N$, hence positive-parity $\Delta^*$'s with $L$ even and
$N=0$ are mass degenerate with negative-parity $\Delta^*$'s with 
orbital angular momentum $L-1$ and $N=1$. In absence of
spin-orbit forces, the four positive-parity  $\Delta^*$'s with
$J = L-3/2,...,L+3/2$ have the same mass, and so have the 
four negative-parity states. But the $L$ values differ by one unit,
the quartet of positive-parity  $\Delta^*$'s is shifted to the
right. Only six states form parity doublets, two states remain
'solitaires', the negative-parity state with $J=L-1-3/2$ and the
positive-parity state with $J=L+3/2$. This effect is visualized
in Fig. \ref{n-delta} for $(L+N,P) = 2\pm,\,4\pm,\,6\pm$. The solitaire
states are separated from their parity partners by one spacing
$a=1.142$\,GeV$^2$. The spacing is even larger ($2a$) when high-mass
$\Delta^*$ resonances all have intrinsic spin 3/2 as we suggested
above.  
\par
We predict that the
$\Delta_{11/2^+}(2420)$ and $\Delta_{15/2^+}(2950)$ should
remain solitaires, should not have close-by chiral partners, in
contrast to the prediction of \cite{Cohen:2001gb}. On the contrary,
a $\Delta_{13/2^+}$ should exist at about 2893\,MeV, about
mass-degenerate with the $\Delta_{13/2^-}(2750)$, in this case in
agreement with the prediction of \cite{Cohen:2001gb}. 
\par
The nucleon mass spectrum is more complicated, as shown in
Fig. \ref{n-delta} (right).  Nearly mass-degenerate
chiral doublets develop due to the $I_{sym}$ term in
eq.~(\ref{mass}). For positive-parity baryons with spin $S=1/2,
I_{sym}= 1$; for negative parity baryons with $S=3/2, I_{sym}=0$.
A positive-parity baryon with orbital angular momentum $L$
thus undergoes a shift downwards (in mass square equal to
the $\Delta$-N mass separation) and is thus found at a mass
not too far from negative-parity N$^*$ resonances having
orbital angular momentum $L-1$ and $S=3/2$. Mass degeneracy is
thus expected but only approximately. The predicted mass splitting 
is small enough that data may mimic parity doublets. Striking
differences are only expected for negative-parity states with
$J=L-3/2$. These states are difficult to establish experimentally but
they are true solitaires. 
\par
A decision if nucleonic resonances form
parity doublets requires a quantitative analysis which is presented next.
First we notice that according to eq.~(\ref{mass}) the
mass difference between two resonances with consecutive $L$   and otherwise
identical quantum numbers vanishes asymptotically:
$M_{L+1}^2 - M_{L}^2 = (M_{L+1} - M_{L})(M_{L+1} + M_{L}) = a$
and hence $M_{L+1} - M_{L} = a/(M_{L+1}+M_{L})$. Asymptotically, all mass
separations vanish with $1/M$ and chiral
symmetry is trivially restored.
\par
We now look for an effect of chiral symmetry beyond this trivial
asymptotic behavior. We do so by comparing the consistency of the data
with the assumption of parity doublets and, alternatively, with
their consistency with  $(L,S)$ multiplets with vanishing $\vec
L\cdot\vec S$ coupling. 
\par
First we calculate the mean mass deviation of baryon resonances
when they are interpreted as parity doublets: 
\begin{equation*}
\qquad\qquad \qquad\qquad \sigma_{parity\ doublets} =
\sqrt{\frac{1}{10}\sum_{i=1,20}(M_{i} - M_{\pm})^2} = 97\ {\rm MeV}
\qquad\qquad \qquad \qquad\qquad  \hfill (3) 
\label{doub}
\end{equation*}
where $M_{\pm}$ are the mean masses of positive- and
negative-parity resonances paired to one parity doublet
(see Table~\ref{chiral}). The sum extends over 20 resonances;
there are 10 degrees of freedom.
\par
We now determine the deviation of baryon masses
from the mean value of a $(L,S)$-multiplet: 
\begin{equation*}
\qquad\qquad \qquad\qquad \sigma_{spin\ multiplets} =
\sqrt{\frac{1}{13}\sum_{i=1,20} (M_i - M_{cg})^2} = 39\ {\rm MeV}
\qquad\qquad \qquad\qquad\qquad   \hfill (4) 
\label{mult}
\end{equation*}
where the $M_{cg}$ are the mean values (center of gravity) of
the 7 multiplets involved. 13 is number of degrees of freedom.
\par
The comparison of the two hypotheses reveals that evidence for
parity doublets in the high-mass spectrum is weak, at most. The data
are better described in terms of $(L,S)$-multiplets embracing
SU(6) multiplets of different $J$ but having the same intrinsic orbital
and spin angular momenta. The symmetry leading to parity doublets is
the vanishing of spin-orbit forces and not a phase transition to
chiral dynamics. 
\par
Finally we examine the possibility that chiral symmetry 
is not yet fully restored but does already influence the mass
spectrum. We do so by testing the hypothesis that the solitaire states
could be slightly 'attracted' by its nearest chiral partner (even
thought the solitaire state remains within its $(L,S)$ multiplet). 
\par 
Indeed, the 
mass of the $\Delta_{7/2^+}$(1950) is larger than the mean of its
three partners of lower $J$, possibly it is 'attracted' by the
$\Delta_{7/2^-}$(2200). The same effect is found for the $L=2$ states
N$_{5/2^+}$(2000)and N$_{7/2^+}$(1990) having masses which are
larger than the N$_{3/2^+}$(1900) and thus closer to the masses
of the N$_{5/2^-}$(2200) and N$_{7/2^-}$(2190) (which have $L=3$). The
N$_{9/2^-}$(2250) (with an assigned $L=3$) is even slightly above the
($L=4$) N$_{9/2^+}$(2220). (In this case, we do not have a neighbor
state to quantify an attraction). If we normalize for these resonances
the mass difference to be zero at the masses at the center of gravity
of a multiplet (1921\,MeV for the four positive parity $\Delta^*$ with $L=2$)
and 1 at the mass of the chiral partner (2200\,MeV),
we find a attraction factor $\gamma_{attr}$ of the $\Delta_{7/2^+}$(1950) of
$\gamma_{attr}=0.10\pm 0.14$. The error is derived assuming 
errors as given in (\ref{mult}). 
The mean attraction factor for the three cases, $\Delta_{7/2^+}$(1950),
N$_{5/2^+}$(2000) and N$_{5/2^+}$(1990), in which $\gamma_{attr}$
can be defined is $\gamma_{attr} =0.13\pm 0.09$. 
There is thus a hint that the solitaire states are attracted by their
parity-doublet partner, even though chiral symmetry breaking effects
still dominate the interaction. Optimistically, the non-zero value
$\gamma_{attr} =0.13\pm 0.09$ can be
seen as onset of a regime in which chiral symmetry is restored.

\section{Conclusions}
We have studied the question if parity doubling observed 
in high-mass N$^*$ and $\Delta^*$ resonances can be interpreted
as evidence for chiral symmetry restoration in baryon excitation. 
We find that 
the appearance of parity doublets does not reflect chiral symmetry 
but rather the vanishing of 
spin-orbit forces in quark-quark interactions in baryons. 
This new interpretation of the parity doublets 
gives predictions for masses of high-mass
baryon resonances which differ distinctively from those based on the
hypothesis of chiral symmetry restoration.
\par
We have searched for indications that chiral symmetry might lead
to a weak attraction between chiral partners. We find
a positive 1.4$\sigma$ effect. Clearly, more precise data are required
to establish an onset of chiral symmetry restoration in the baryon
mass spectrum.

\paragraph{Acknowledgments} I would like to thank S.U. Chung,
L.Y. Glozman, D. Diakonov, U. L\"oring,
B. Metsch, H. Petry, B. Schoch, S.F. Tuan and Chr. Weinheimer
for stimulating discussions on
parity doublets and on the baryon spectrum.
\end{document}